\documentclass[12pt]{iopart}
\usepackage{graphicx}
\newcommand{\asc}{{a_{\rm sc}}}
\newcommand{\aho}{{a_{\rm ho}}}
\newcommand{\abg}{{a_{\rm bg}}}
\newcommand{\ee}{{\rm e}}
\newcommand{\rr}{{\bf r}}
\newcommand{\FF}{\,_1{\rm F}_1}
\begin{document}

\title{
A Two-Atom Picture of Coherent Atom-Molecule Quantum Beats
}

\author{Bogdan Borca\dag, D. Blume\ddag, Chris H. Greene\dag}

\address{\dag\
JILA and Department of Physics,
University of Colorado, Boulder, Colorado 80309-0440 }
\address{\ddag\
Department of Physics,
Washington State University, Pullman, Washington 99164-2814}
\ead{borca@jilau1.colorado.edu}
\ead{doerte@wsu.edu}
\ead{Chris.Greene@Colorado.edu}

\begin{abstract}
A simple two-atom model is shown to describe
a Bose-Einstein condensate of alkali atoms
subjected to external magnetic 
field ramps near a Feshbach resonance. The implications
uncovered for two atoms in a trap can be applied at least
approximately to a many-atom condensate.  A connection
to observations is accomplished by scaling the 
trap frequency to achieve a density comparable to that of the
experiments, which yields the fraction
of atom pairs in the gas that become molecules.  
A sudden approximation is used to model
the external magnetic field ramps in the vicinity of a 
two-body Feshbach resonance. 
The results of this model are compared 
with recent experimental 
observations of Donley {\it et al.} \cite{Donley02}.
\end{abstract}


\submitto{\NJP}

\maketitle

\section{Introduction}
Utilizing Feshbach resonance physics,
recent experiments have produced an
atomic Bose-Einstein condensate (BEC) coherently
coupled to molecules in high vibrationally excited bound states
\cite{Donley02}. 
The coupling of atomic and molecular states
was achieved through application of a pulsed external
magnetic field and it has sparked much 
interest \cite{Zoller02} since
it ultimately (although possibly not yet) 
should lead to the creation of a
molecular condensate.
The problem of creating a molecular BEC
has been one of the focus areas of 
ultracold physics research for several years now
\cite{Heinzen00}. Two different
techniques, namely photoassociation and a 
magnetic field ramp near a Feshbach resonance, 
have been used in attempts to transform
an atomic condensate into a molecular one.
The use of Feshbach resonances to control 
the atom-atom scattering length, and other properties,
has been previously demonstrated 
experimentally \cite{Inouye98, Wieman00}. 

It was predicted theoretically \cite{Timmermans99} that 
magnetic field pulses would drive a significant part of 
an atomic BEC into a molecular one. 
These predictions were based on a mean field theory  
approach, of the same 
type that has proven very successful in 
describing many properties of the alkali atom BECs produced experimentally
to date. However, two sets of experiments performed
at JILA have shown puzzling results that 
did not match the original theoretical predictions.
In one experiment, a single magnetic pulse close to the Feshbach
resonance probed the strongly interacting atomic
dynamics, \cite{Claussen02} while in a second experiment
double pulses generated
interference patterns between the different states
populated. In both experiments, the 
atoms, part of the initial BEC cloud, were observed
to end up in one of the following three components: a remnant BEC cloud,
a burst of hot atoms, and a missing (undetected) component.
This outcome differed from the theoretical predictions
which only accounted directly for two components. The questions 
regarding the nature of the three components observed experimentally, 
including the specific issue of whether a molecular BEC was created, 
were addressed by two independent  theoretical papers
\cite{Kokel02,Kohler03}.  Both of these  papers 
accounted for many body effects by employing field theory beyond 
the mean field (i.e., Gross-Pitaevskii equation) level.
These two studies gave similar answers, in identifying the observed 
atom bursts as hot, non-condensed atoms, and the missing 
component as a molecular condensate. 
In addition to Refs.  \cite{Kokel02} and \cite{Kohler03}, three 
other theoretical studies involving
many-body approaches have also addressed the interpretation
of the Donley {\it et al.} experiment
\cite{Braaten03,Mackie02,Stoof03}. 

In this paper we propose an alternative model 
that does not rely on field theory.
Our model considers only two-body dynamics
and uses a very simple scaling procedure 
to apply our results
to the many-body system studied experimentally.
The observed oscillatory behavior can then be viewed
as a simple example of quantum beats of the type that
arise whenever indistinguishable quantum mechanical pathways
associated with two or more impulsively-excited stationary 
states interfere coherently.\cite{Andra1970}
Specifically, our treatment considers
two atoms confined by 
a spherically symmetric harmonic oscillator potential,
which interact through a contact potential.
The stationary
states of such a system have been described by
Busch {\it et al.}, \cite{Rzazewski98} who proposed
it as a model that can be used in the context
of ultracold collisions occuring in trapped gases.
The properties of this treatment were further investigated
in the context of the condensed Bose gases by 
Julienne and coworkers \cite{Tiesinga00,Bolda02} and by
Blume and Greene \cite{Blume02}.
These papers investigated the strong  
interactions that occur near a Feshbach resonance,
and employed an energy-dependent scattering length
to model this situation. 
Here we implement this model for a time-dependent magnetic field, 
augmented by the sudden approximation to model the rapid field ramps.

By performing a frequency rescaling, an initial condensate 
density can be achieved for two bodies 
that is comparable to the density range studied experimentally. 
The premise of our model is that the energy scale of the trap
energy levels is very low in present day experiments, 
far smaller than the molecular binding energies of interest.
At the same time, we anticipate that the 
physics of any single molecule
formation is controlled by the interaction of just two atoms, 
even in a many-atom condensate.  Accordingly, we consider just 
two atoms in an oscillator trap of very high frequency, adjusted
so that the density of the two atoms
becomes  the same as  the condensate density in the experiments.
The resulting approach is then used to model the recent experimental results
of Donley {\it et al.} \cite{Donley02}.  Our two-body model is
shown to describe  most of the nontrivial features observed in the
experiment, although some discrepancies remain.  
This may indicate that a more elaborate inclusion of the many-body effects
may be necessary to achieve a full quantitative description.  
Nevertheless, our results show that more of the key  effects
can be interpreted in terms of two-body physics
than appears to have been realized in existing theoretical models.

This paper is organized as follows: 
Section 2 determines the eigenstates 
of two trapped atoms using a quantum-defect-style
method that differs from the treatment of Ref.
\cite{Rzazewski98} but is equivalent.
Section 3 discusses the behavior of the atom pair 
close to 
a Feshbach resonance, and our
approximate solution of the time-dependent 
Schr\"odinger equation using
the sudden approximation. Section 4
discusses the frequency rescaling employed
to interpret the many-atom system.  Section 5
compares the results of our model with recent experiments,
while Section 6 summarizes our conclusions.

\section{Two Interacting Atoms in a Trap}

We consider two atoms of mass $m$ in a spherical oscillator trap 
of angular frequency $\omega$, which interact through a 
zero-range potential $V(\rr)$ \cite{zrpbook},
\begin{equation}
V(\rr)=\frac{4 \pi \hbar^2 \asc}{m}
\delta^{(3)}(\rr)
\frac{\partial}{\partial r}r.
\label{zrp}
\end{equation}
Here $\asc$ is the two-body atom-atom
scattering length and $\rr$ is the relative coordinate
of the two particles. The Hamiltonian of the two-body system separates 
into a center of mass part and a relative part. 
The center of mass part has the usual
harmonic oscillator solutions which are not affected by
the scattering length; hence, we  
focus on the relative motion in the following.
 
Since the contact potential in
Eq. (\ref{zrp}) acts solely on the $s$-wave symmetry, 
we consider only 
solutions of the relative Schr\"odinger equation with zero orbital
angular momentum. We define the harmonic oscillator length 
$\aho=\sqrt{\hbar/(\omega m/2)}$ corresponding to the
{\it reduced } mass, $m/2$, as our length scale, 
leading to a dimensionless radial coordinate
$x=r/\aho$. 
Our energy unit is chosen to be $\hbar \omega$,
resulting in a dimensionless energy variable  
$\epsilon=E/\hbar \omega$. 
The $s$-wave eigenfunction $\phi_{\epsilon,l=0}$ 
of the radial Schr\"odinger equation in the relative
coordinate corresponding to energy $\epsilon \hbar \omega $
is then rescaled,
\begin{equation}
\phi_{\epsilon,l=0}=
\frac{u(x)}{x} \frac{1}{\sqrt{4 \pi}}\,,
\label{hoeq}
\end{equation}
so that the radial
equation has only second derivatives, 
\begin{equation}
\left( 
-\frac{1}{2}\frac{d^2}{dx^2} + \frac{1}{2}x^2 
\right) u(x)=\epsilon\,u(x)\,.
\label{radeq}
\end{equation}
The contact potential, Eq. (\ref{zrp}),
imposes a boundary condition on the logarithmic derivative
of $u(x)$ at the origin:
\begin{equation}
\frac{u'(0)}{u(0)}=-\frac{\aho}{\asc}\,.
\label{boundaryatorigin}
\end{equation}
Solutions of these equations have been obtained 
by Busch {\it et al.} \cite{Rzazewski98}. However, 
we now rederive these
solutions in a slightly different manner, along the lines
of quantum defect theory (QDT) \cite{Greene79}.

We start with a pair of solutions of Eq. (\ref{radeq}), $f$ and $g$,
that have regular,
\begin{equation}
f_\nu(x)=A_{\nu}\, x \ee^{-x^2/2} \FF(-\nu;\frac{3}{2};x^2),
\label{fregular}
\end{equation}
and irregular,
\begin{equation}
g_\nu(x)=B_{\nu}\, \ee^{-x^2/2} \FF(-\nu-\frac{1}{2};\frac{1}{2};x^2),
\label{girregular}
\end{equation}
behavior at the origin, at any energy.   Here 
$\nu$ denotes a quantum number, $\nu=\epsilon/2-3/4$,
while $A_{\nu}$ and $B_{\nu}$ are constants that 
will be determined later. In the following, 
our solutions are characterized by the subscript
$\nu$ instead of the subscript $\epsilon$.
We calculate the asymptotic behavior of the two solutions using 
the known behavior of the confluent hypergeometric
function, $\FF$,
\begin{equation}
\left.\FF(a;b;x)\right|_{x\to \infty} 
\rightarrow 
\frac{\Gamma(b)}{\Gamma(a)} x^{a-b} \ee^{x} \, +
\,\cos(\pi a) \frac{\Gamma(b)}{\Gamma(b-a)} x^{-a} \, ,
\end{equation}
as well as the gamma reflection formula:
\begin{equation}
\Gamma(\nu)\,\Gamma(1-\nu)=\frac{\pi}{\sin (\pi \nu)}.
\end{equation}
For $x\to\infty$, we obtain 
\begin{eqnarray*}
\fl
f_{\nu}\rightarrow A_\nu\Gamma(\frac{3}{2})
\left(
-x^{-2\nu-2} \ee^{x^2/2}  \sin(\pi \nu) \frac{\Gamma(\nu+1)}{\pi}
\, + \, 
x^{2\nu+1} \ee^{-x^2/2}\cos(\pi \nu)\frac{1}{\Gamma(\nu+\frac{3}{2})}
\right),
\\
\fl
g_{\nu}\rightarrow B_\nu\Gamma(\frac{1}{2})
\left(
-x^{-2\nu-2} \ee^{x^2/2}  \cos(\pi \nu) \frac{\Gamma(\nu+\frac{3}{2})}{\pi}
\, -\, 
x^{2\nu+1} \ee^{-x^2/2}\sin(\pi \nu)\frac{1}{\Gamma(\nu+1)}
\right)\,.
\end{eqnarray*}
Our goal is to recast these solutions in the ``usual''
QDT form \cite{Greene79} given by
\begin{eqnarray}
f_{\nu}\rightarrow \,\,\,-C 
\left( D^{-1} \ee^{x^2/2} x^{-2\nu-2} \sin(\pi \nu) 
\,-\, D\, \ee^{-x^2/2} x^{2\nu+1} \cos(\pi \nu) \right),
\label{fgcd_a}
\\
g_{\nu}\rightarrow C
\left( D^{-1} \ee^{x^2/2} x^{-2\nu-2} \cos(\pi \nu)
\,-\, D\, \ee^{-x^2/2} x^{2\nu+1} \sin(\pi \nu) \right)\,.
\label{fgcd_b}
\end{eqnarray}
In addition, we want to normalize the functions $f$ and $g$ such
that their Wronskian ${\rm W}[f_\nu , g_\nu]$ is $2/\pi$.
These requirements can be fulfilled by defining $A_\nu$
and $B_\nu$ [Eqs. (\ref{fregular}) and (\ref{girregular}),
respectively] appropriately, which leads to 
\begin{eqnarray}
f_\nu=\frac{2}{\sqrt{\pi}}
\sqrt{\frac{\Gamma(\nu+{3\over 2})}{\Gamma(\nu+1)}}
\,x\,\ee^{-x^2/2}\,\FF(-\nu;\frac{3}{2};x^2)\,,
\label{fnu}
\\
g_\nu=-\frac{1}{\sqrt{\pi}}
\sqrt{\frac{\Gamma(\nu+1)}{\Gamma(\nu+{3 \over 2})}}
\,\ee^{-x^2/2}\,\FF(-\nu-\frac{1}{2};\frac{1}{2};x^2)\,.
\label{gnu}
\end{eqnarray}
Using these normalizations, the 
constants $C$ and $D$ [Eqs. (\ref{fgcd_a}) and (\ref{fgcd_b})]
become
$C=1/\sqrt{\pi}$ and $D=\sqrt{\pi}/
\sqrt{\Gamma(\nu+1)\Gamma(\nu+3/2)}$.

With these solutions for $f$ and $g$ in hand, we can proceed
in the spirit of QDT by  deriving a solution to the 
radial Schr\"odinger equation that accounts for
an additional non-oscillator short-range 
potential.  For distances beyond those
where the short-range potential is non-negligible, the 
radial wave function $u_\nu$ must assume the form 
\begin{equation}
u_\nu\,=\,f_\nu\,\cos\pi \mu \,-\,g_\nu\,\sin\pi \mu \,.
\label{qdteq}
\end{equation}
Armed with the known asymptotic behavior of $f$ and $g$, we can 
determine the asymptotic behavior of $u$ 
and impose the requirement 
that $u$ is finite at large $x$
(i.e., the coefficient of the growing exponential is zero).
This leads to the equality $\sin\pi (\nu+\mu)=0$, which 
can be recast as a quantization condition
\begin{equation}
\epsilon=2 (n-\mu) +\frac{3}{2}\,,
\label{queq}
\end{equation}
where $n=0,1,2,...\,$.  
The last step towards finding the energy levels $\epsilon$ of our
confined atom pair is to impose the boundary condition on the
solution $u$, which is implied by the
contact potential at the origin [Eq. (\ref{boundaryatorigin})]. 
Using the fact that at small arguments $\FF(a;b;x)$
approaches $1$ \cite{Abramowitz}, we obtain
\begin{equation}
\frac{u'(0)}{u(0)}=\,-
\frac{f'(0) \cos\pi \mu }{g(0) \sin \pi \mu}
\,=\,
-\frac{\aho}{\asc}\,. 
\end{equation}
Upon inserting the explicit forms for $f$ and $g$ 
as given in Eqs. (\ref{fnu}) and (\ref{gnu}),
we obtain  the following 
equation for the quantum defect $\mu$ (see also \cite{Blume02})
\begin{equation}
\tan \pi \mu =-\frac {\asc}{\aho} 
\frac{2\,\Gamma\left( {\epsilon \over 2} + {3 \over 4} \right)}
{\Gamma\left( {\epsilon \over 2} + {1 \over 4}\right)}\,.
\label{queq2}
\end{equation}
Equations (\ref{queq}) and  (\ref{queq2}) allow 
determination of the energy spectrum for any value of
the scattering length $\asc$. It can be shown that these
equations are equivalent to the transcendental
equation 
\begin{equation}
\frac{2 \Gamma \left(- {\epsilon \over 2} + {3 \over 4} \right)
}{\Gamma \left( -{\epsilon \over 2} + {1 \over 4}\right) }
=\frac{1}{\asc/\aho} \,
\label{treq}
\end{equation}
of Busch {\it et al.} \cite{Rzazewski98}.
The energy quantization conditions derived here for
a zero-range pseudo potential using QDT also apply to a confined atom pair
interacting through an arbitrary (i.e., non-contact) short-range potential,
the only difference being that the quantum defect $\mu$ has a 
different value.
Regardless of the specific short-range potential, 
the corresponding eigenfunctions $u_\nu$ are 
then given outside the potential range 
in terms of the hypergeometric U function  
\cite{AbramU,Rzazewski98},
\begin{equation*}
u_{\nu}(x)=N_{\nu} {\rm e}^{-x^2/2} {\rm U}
\left(\frac{-(2 \nu+1)}{2},\frac{1}{2},x^2 \right),
\label{wfeq}
\end{equation*}
where $N_{\nu}$ is a normalization constant.


\section{Two atoms near a Feshbach Resonance. Overlap matrix elements}
\label{overlapmatrixelementsection}

To apply our formalism derived above to
two-atom states that lie energetically near a Feshbach resonance,
we rewrite our scattering length $\asc$ as a function of the magnetic
field strength $B$
\begin{equation*}
\asc(B)=\abg\left(1- \frac{\Delta}{B-B_0} \right).
\label{asc}
\end{equation*}
Here, $\abg$ denotes the scattering length far from the resonance, 
$B_0$ denotes the resonance position, while $\Delta$ is a 
parameter related to the width of the resonance.
In our numerical calculations (see Sec. \ref{numericalresults}),
we choose parameter values  in agreement with 
those obtained by fitting to recent  experimental data
\cite{Claussen03} of the Feshbach 
resonance in $^{85}$Rb:  $B_0=155.041$ G,  $\Delta=10.71$ G, 
and $\abg=-443$ a.u. 
Note that a  similar model was employed in Ref. \cite{Bolda02} 
for the case of a Feshbach
resonance in $^{23}$Na. While that study additionally
incorporated an energy dependence of the scattering length,
the energy dependence of $\asc(B)$ in the present study
is comparatively weak over the range of energies and magnetic fields 
considered. Hence  we choose to neglect it.

Our goal is to describe
our two-atom system when it is subjected to 
time-dependent magnetic field ramps $B(t)$, 
like the ones used in the experiments of \cite{Donley02}.
To approximate the experimental $B(t)$,
we assume a piecewise constant $B(t)$ as indicated
in Fig. \ref{fig3}.
In our approach
the assumption of an instantaneous variation of $B(t)$
amounts to the use of the sudden approximation.
In this approximation, the wavefunction of the atom pair 
is initially unaffected by the instantaneous change of $B(t)$. 
Let the magnetic field strength before (after) the instantaneous
change be $B_1$ ($B_2$).
It is then convenient to write the time-dependent superposition
state $\Psi$ at field strength $B_1$ in terms of the
eigenfunctions
$\{\phi_{\nu_1}\}_{\nu_1=0,\infty}$ 
(with corresponding eigenvalues 
$\{\epsilon_{\nu_1}\}_{\nu_1=0,\infty}$), 
and that at field strength $B_2$ in terms of the
eigenfunctions
$\{\phi_{\nu_2}\}_{\nu_2=0,\infty}$ 
(with corresponding eigenvalues 
$\{\epsilon_{\nu_2}\}_{\nu_2=0,\infty}$),
\begin{equation}
\Psi(B_j,t)=\sum_{\nu_j} a^{(j)}_{\nu_j}(t) \phi_{\nu_j}\,
,\,\,\,j=1,2\,.
\end{equation} 
The ``new'' expansion coefficients $a^{(2)}_{\nu_2}$ can 
be expressed through the ``old'' expansion coefficients,
\begin{equation*}
a^{(2)}_{\nu_2}=\sum_{\nu_1}O_{\nu_2,\nu_1}\,a^{(1)}_{\nu_1}\,,
\end{equation*}
where $O_{\nu_2,\nu_1}$ denotes the overlap matrix 
between the eigenfunctions corresponding to $B_1$ and $B_2$,
respectively,
\begin{equation}
O_{\nu_2,\nu_1}= \left\langle  \phi_{\nu_2} |\phi_{\nu_1} \right \rangle \,.
\label{overlapmatrix}
\end{equation}

\begin{figure}
\vspace{10 mm}
\begin{center}
\includegraphics[width=4in]{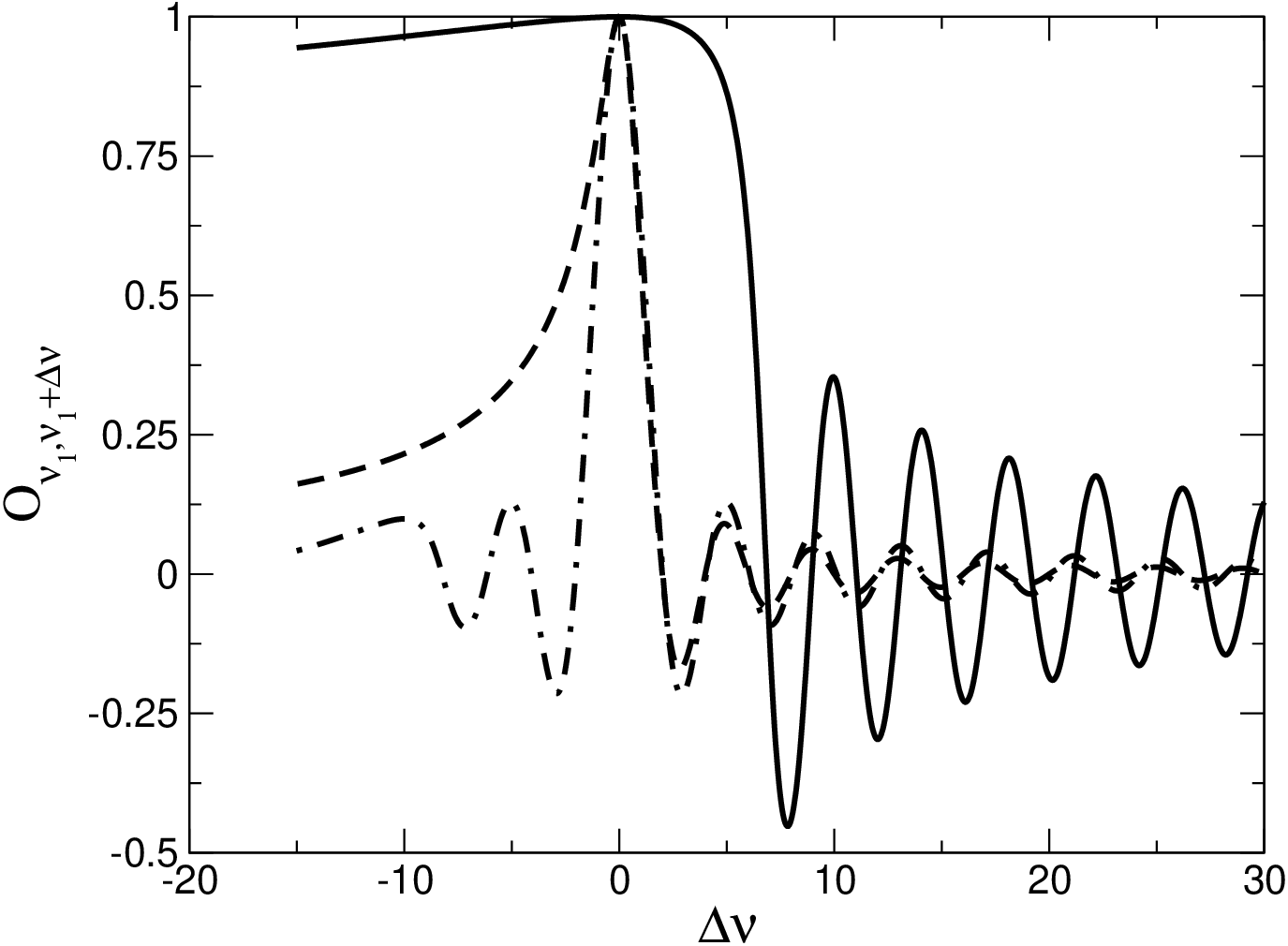}
\end{center}
\caption{\label{fig1}
Dependence of the overlap matrix element $O_{\nu_2, \nu_1}$
between two eigenfunctions
corresponding to quantum numbers $\nu_1$ and $\nu_2=\nu_1+\Delta \nu$ 
on the quantum number difference, $\Delta \nu$, for three
values of $\nu_1$: $\nu_1=-5$ (solid line), 
 $\nu_1=1.5$ (dashed line), and $\nu_1=10$ (dashed-dotted line).
Note that the solid line in this figure represents the 
projections of the molecular state onto the other trapped atom states.
}

\end{figure}

The following derivation shows how the 
eigenstate transformation projections $O_{\nu_2,\nu_1}$ can be determined
analytically.  Let $u_{\nu_{1,2}}$ 
be the reduced radial wavefunctions corresponding to $\phi_{\nu_{1,2}}$. 
If we multiply the equation for $u_{\nu_{1}}$ by $u_{\nu_{2}}$
and that for  $u_{\nu_{2}}$ by $u_{\nu_{1}}$, subtract the resulting equations
and integrate the result from 0 to $\infty$, an integration
by parts gives
\begin{equation}
\left(\epsilon_{\nu_{1}}-\epsilon_{\nu_{2}}\right)
\int_0^\infty u_{\nu_{1}}(x) u_{\nu_{2}}(x) dx =W\left[u_{\nu_{1}},
u_{\nu_{2}}\right]_{x=0}\,.
\end{equation}
Here, $W\left[u_{\nu_{1}},
u_{\nu_{2}}\right]_{x=0}$ denotes
the Wronskian of $u_{\nu_{1}}$
and $u_{\nu_{2}}$
evaluated at the origin. The above formula allows the
determination of both the normalization constant $N_\nu$ 
[Eq. (\ref{hoeq})] and the
overlap matrix element $O_{\nu_2,\nu_1}$ [Eq. (\ref{overlapmatrix})], 
\begin{eqnarray}
\fl
O_{\nu_2,\nu_1}=N_{\nu_1}\,N_{\nu_2}&
\left[
2\,\cos (\pi \,\nu_1 )\,\Gamma(\frac{3}{2} + \nu_1 )\,
\Gamma(1 + \nu_2 )\,\sin (\pi \,\nu_2)
\right.
\nonumber
\\
\fl
&-2\,
\left. \cos (\pi \,\nu_2)\,\Gamma(1 + \nu_1 )\,
\Gamma(\frac{3}{2} + \nu_2 )\,\sin (\pi \,\nu_1) 
\right]/
\left[ 4\, \pi\,
\left( \nu_2  - \nu_1  \right) \right]
\label{o12}
\end{eqnarray}
and
\begin{equation}
\fl
N_{\nu}=\left\{
\Gamma(1 + \nu )\,\Gamma(\frac{3}{2} + \nu )\,
\left[ 2\,\pi  + \left( \psi(\nu) -\psi(\frac{1}{2} + \nu)  \right) 
\,\sin (2\,\pi \,\nu ) \right]
\right\}^{-1/2}\,,
\end{equation}
where $ \psi$ denotes the digamma function \cite{Abramowitz}.
The overlap matrix elements $O_{\nu_2, \nu_1}$ are expressed 
in terms  of the $\nu$ quantum numbers,
which, in turn, are directly linked to the energy eigenvalues 
$\epsilon$ through $\epsilon=2\nu+3/2$. 
Figure \ref{fig1} 
shows examples of the dependence of the overlap matrix 
elements on the non-integer quantum numbers $\nu_1$ and $\nu_2$.
While the overlap matrix elements oscillate rapidly
with $\Delta \nu$ ($\Delta {\nu}=\nu_2 -\nu_1$),
our analytical formula for the matrix elements results
in stable numerical calculations (see Sec. \ref{numericalresults}).

\section{Description of many-body effects through a frequency rescaling}

The simple model described in the previous section allows
us to describe the states of a trapped atom
pair that undergoes sudden changes of the interatomic interaction,
here parameterized accurately through a magnetic field-dependent 
scattering length.
Our goal is now to apply our two-atom model to 
interpret an ensemble of 
$N$ atoms ($N \gg 2$).
Without constructing a rigorous many-body approach like, e.g., 
the ones based on the field theory formalism, we will 
attempt to account for many-body effects by 
using our two-particle model with 
a rescaled frequency. The idea suggested by 
Cornell \cite{Cornell2002} is 
to capitalize on the importance of the diluteness or gas
parameter $n\,\asc^3$, ($n$ is the density of atoms)
which in more rigorous many-body models of 
degenerate Bose gases plays a key role in 
determining the behavior of the system. 
Instead of modeling an $N$-atom system we will 
model a two-body system that has the same diluteness
parameter as the experimentally studied 
$N$ atom sample.

To motivate that our two-body description can,
at least to a crude level of approximation,
account for many-body physics
we calculate the overlap integral for one particular case.
We consider the overlap integral between a two-body  state
located very far from the resonance centered at $B_0$,
$u_{\rm ho}(r)$, with that corresponding
to a value of $B$ close to $B_0$, $u_M(r)$.
$u_{\rm ho}$ refers to the trap ground state, which we approximate 
through a state describing two independent atoms with zero
scattering length,
\begin{equation}
u_{\rm ho}(x)\approx\frac{2}{\pi^{1/4} }x {\rm e}^
{-x^2/2}
\end{equation}
$u_M(r)$ denotes
a molecular state (i.e., the state that remains bound even in the 
absence of the confining potential). Neglecting the
influence of the confining potential we assume the wavefunction
of this state to be:
\begin{equation}
u_M(r)\approx\sqrt{2 \kappa}{\rm e}^{-\kappa r}\,,
\end{equation}
where $\kappa=1/\asc$. We estimate the overlap integral of these  two states,
by assuming that the exponential in $u_{\rm ho}$ is approximately 1 
over the range relevant for  the evluation of the integral. 
This aproximation yields
\begin{equation}
\left\langle u_M | u_{\rm ho} \right \rangle \approx
\left( {2 \asc \over \aho} \right)^{3/2} \, \pi ^{-1/4}.
\end{equation}
The absolute value of this overlap matrix element gives
the probability 
$p$ that the initially ``unbound'' atom pair
ends up in the molecular state as the $B$ field
is tuned close to resonance,
\begin{equation}
p=\frac{8}{\sqrt{\pi}} \left( \frac{\asc}{\aho}\right)^3\,.
\end{equation}
For small $p$ ($p \ll 1/N$), we can extend our two-body treatment to model
an ensemble of $N$ atoms.
As the $B$ field is tuned to resonance,
{\em{each atom pair}} in the ensemble has the probability $p$ to
form a molecular bound state.
Using this simple picture, the calculation of the fraction of atoms that
transform into molecules,
$f_{atoms \rightarrow molecules}$, due to the
magnetic
field ramp amounts to a simple counting of atom pairs,
\begin{equation}
f_{{\rm atoms} \rightarrow{\rm{ molecules}}} =\frac{2}{N}\frac{N(N-1)}{2}p
\approx Np = \frac{8}{\sqrt{\pi}}  \frac{\asc^3 N}{\aho^3}\,.
\end{equation}
Within our estimate, the fraction of atoms transformed to 
molecules is proportional to the diluteness parameter $n\,\asc^3$ 
(the density is proportional to $N/\aho^3$).
This back-of-the-envelope
estimate, although not a
rigorous proof, provides a somewhat quantitative motivation 
for the idea that lies behind 
the frequency rescaling introduced above.

Since in our two body model we will 
choose a scattering length that is the same as in the sample 
used in the experiment,
having the same diluteness parameter for the two amounts to having
the same density. This condition, that our two atoms confined
by the harmonic potential have the same peak density 
as the average many-body density, $n$,
provides the criterion for determining our rescaled frequency $\omega'$.
For the purpose of evaluating  $\omega'$ from this condition
we neglect the atom-atom potential  and consider the two 
non-interacting atoms in the ground state of the harmonic trap.  
Based on this assumption we  find the relation between
the density of the many-body sample, $n$,  and the
trap frequency confining the two atoms, $\omega'$, to be
\begin{equation}
\omega'=\frac{\hbar}{m_{\rm atom}} \pi
\left( \frac{n}{2} \right)^{2/3}\,.
\label{omegap}
\end{equation}
The experiment \cite{Donley02}
uses an inhomogeneous trap that has a geometric mean 
of $\omega\approx 2\,\pi\,12$ Hz, and involves approximately $N=17100$ atoms.
The starting value of the magnetic field of about 
$162.2$ G corresponds to a scattering length of 220 a.u..
For these parameters, using either the Thomas-Fermi formula
or the Gross-Pitaevskii equation, one obtains
a density profile with an average of approximately 
$n\approx 4\cdot 10^{12}$ cm$^{-3}$
(see, e.g., Fig. 10 of Ref. \cite{Kohler03}). 
Note that this is lower than the claimed experimental density
of $n=1.1\cdot 10^{13}$ cm$^{-3}$.
We find that the final results (see Fig. \ref{fig4}) 
depend sensitively on density, and this way of rescaling 
the frequency in Eq. (\ref{omegap})
seems to best reproduce the experimental results.
The rescaled frequency 
that we obtain for this density 
($n=4\cdot 10^{12}$ cm$^{-3}$)
 is $\omega' \approx 2\,\pi\,3.72$ kHz.

\begin{figure}
\vspace{10 mm}
\begin{center}
\includegraphics[width=4in]{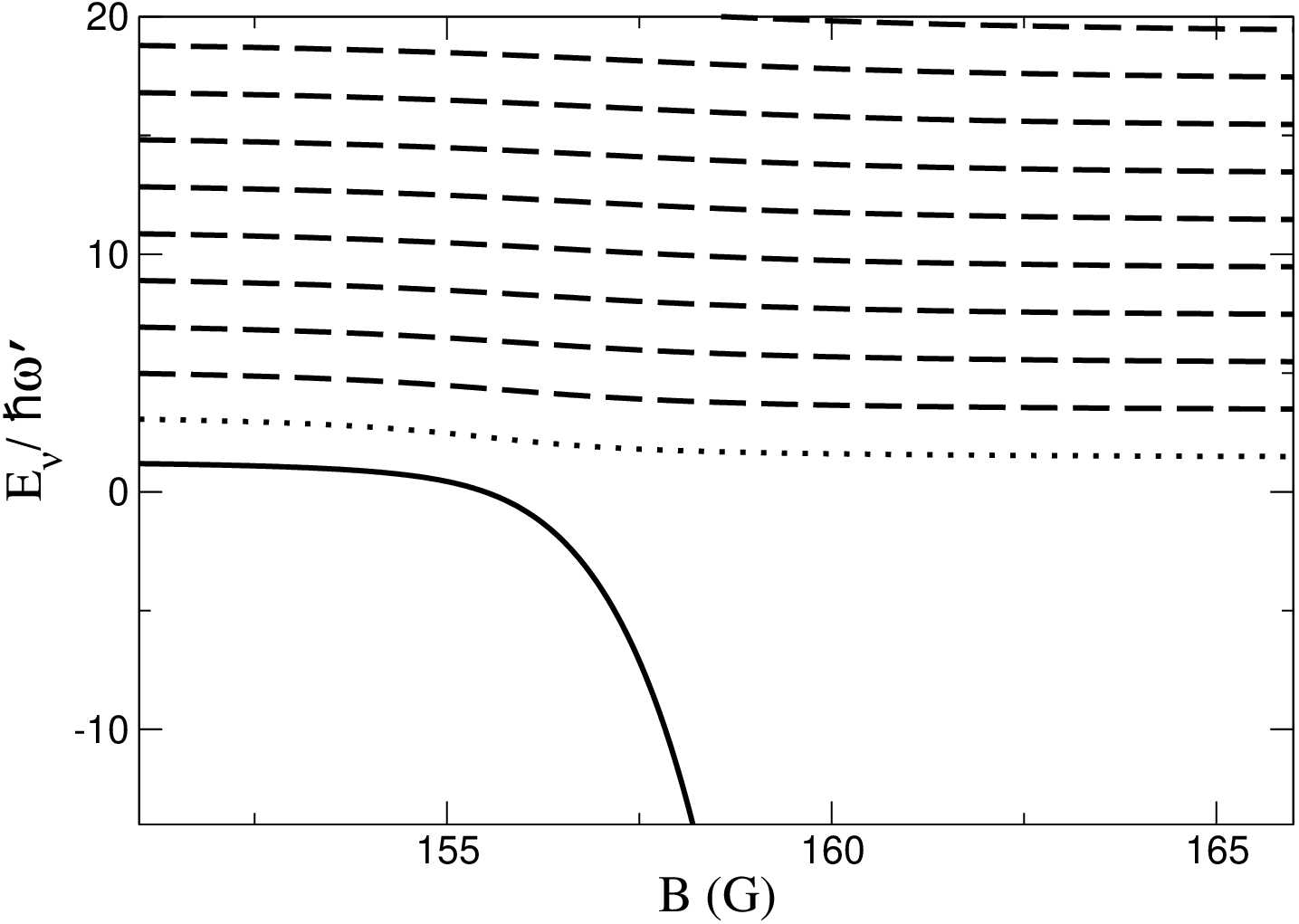}
\end{center}
\caption{\label{fig2}
Energy eigenvalues for two $^{85}$Rb atoms 
for a rescaled frequency $\omega'=2\, \pi \,3.72$ kHz 
(corresponding to $N=17100$) near a Feshbach resonance
centered at $B_0=155.041$ G with $\abg=-443$ a.u. 
and $\Delta=10.71$ G. We distinguish three groups of states:
the molecular state (solid line), the trap ground state
(dotted line), and the trap excited states (dashed lines).
}
\end{figure}

Figure \ref{fig2} shows the eigenvalues
of two $^{85}$Rb atoms confined by a harmonic potential 
with angular
frequency $\omega'$ as a function of the magnetic field strength $B$ in
the vicinity of the Feshbach resonance described through the parameters
given in Sec. \ref{overlapmatrixelementsection}.
Notice that the sequence of avoided crossings of energy levels associated with
the Feshbach resonance becomes smeared out at this high frequency.  
We group the two-body states into three groups (see Fig. \ref{fig2}): 
the molecular state, the trap ground state, and the excited states 
of the trap. 
To connect our two-body study to the $N$-body systems studied 
experimentally, we make the following correspondence 
between the two-atom states and the many body states.
The molecular state of two atoms is in the ground state of the center-of-mass motion, and accordingly we associate this population with translationally cool (condensed)
molecules in the $N$-atom system. At fields above the resonance, the lowest {\it positive} energy state
(which we refer to as the trap ground state) corresponds to condensate
atoms.  Finally,  the higher trap excited states correspond to non-condensed atoms,
i.e., the experimentally observed ``jets of hot atoms'' \cite{Donley02}.
In the next section, we interpret the 
occupation probability of the two-atom
states as the fraction of atoms 
ending up in the corresponding many body states.

\section{Numerical Results}
\label{numericalresults}
\begin{figure}
\vspace{10 mm}
\begin{center}
\includegraphics[width=4in]{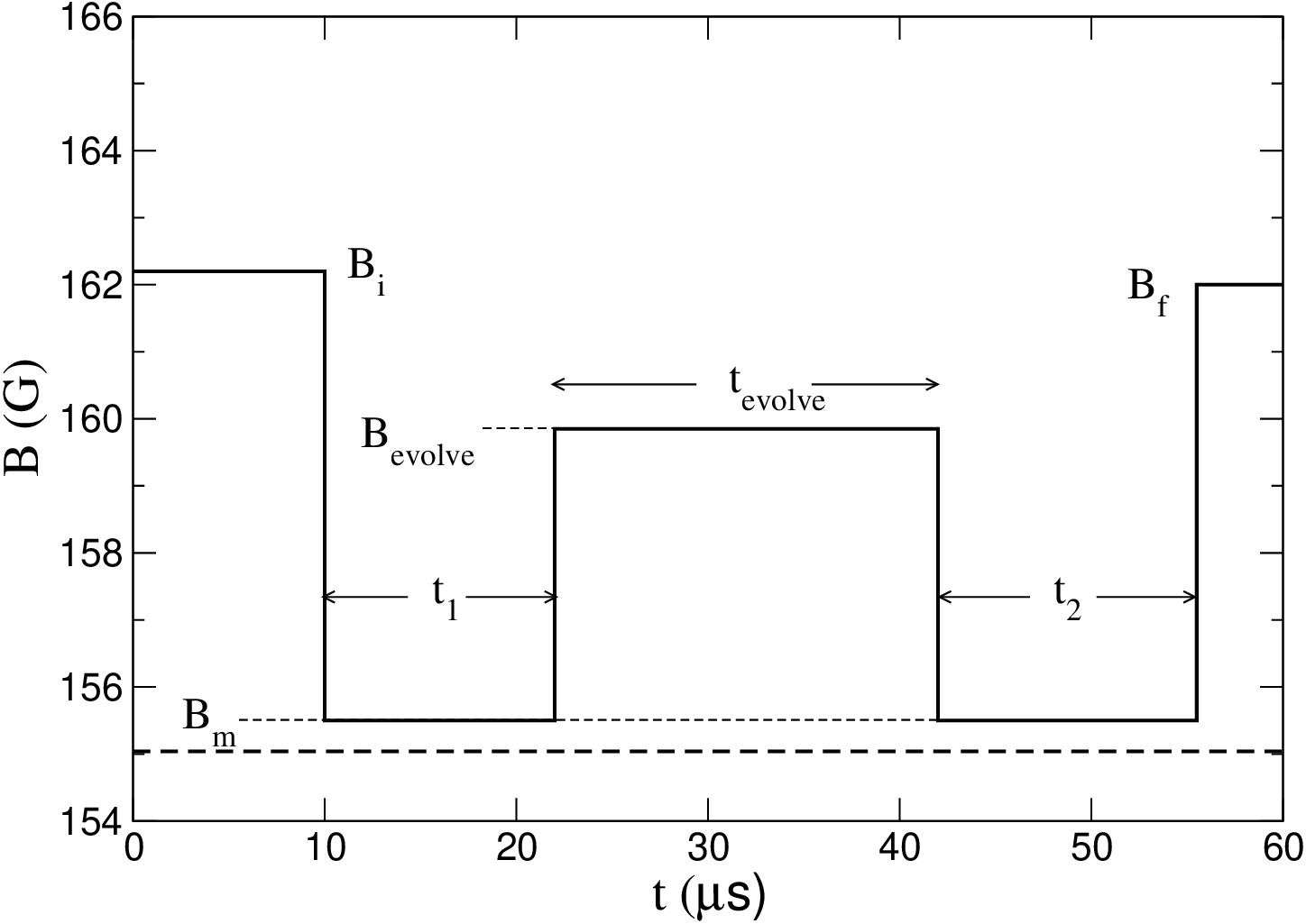}
\end{center}
\caption{\label{fig3}
Magnetic field $B$ as function of time $t$
in our modeling of the experiments of Donley {\it et al.}.
The thick dashed line represents the position of the 
Feshbach resonance, $B_0=155.041$G.
The abrupt variations of the magnetic field
reflect our use of the sudden approximation.
The parameteres  of the pulse are: $B_i=162.2$ G,
$B_m=155.5$ G,  $B_{\rm evolve}=159.85$ G,  $B_f=162$ G,
and $t_1=12$ $\mu$s, $t_2=13.6$ $\mu$s.
}
\end{figure}

We  now use the above description
to simulate the experiment of Donley {\it et al.}
\cite{Donley02}. This experiment consists of applying 
two magnetic field pulses, separated by a time interval
of variable length (denoted by $t_{\rm evolve}$ in Fig. \ref{fig3}),
to a condensed sample of $^{85}$Rb atoms. 
The two pulses of durations $t_1$ and $t_2$, respectively,
ramp the magnetic field to $B_m$, a value 
close to the Feshbach resonance, which couples
the atomic and molecular states of the sample.
After the application of these magnetic pulses, three distinct
components  are  identified experimentally:
a remaining sample of condensed atoms, a hot burst of atoms,
and a ``missing'' component which is not detected under
the current experimental conditions. At least two theoretical 
approaches \cite{Kokel02,Kohler03}, both  involving a
field theory formalism, identified this third component 
(i.e., the missing component) as a molecular condensate,
however, other interpretations exist. 
The size of the three components oscillates
as a function of the evolve time $t_{evolve}$
with a frequency that corresponds to the energy of 
a weakly-bound, vibrationally excited state
of the $^{85}$Rb$_2$ dimer. 

We model the experimental situation
using the sudden approximation to describe
the sharp rises and drops of the magnetic field.
At time $t=0$ in Fig. \ref{fig3}, our initial state is 
chosen to be the trap ground state 
corresponding to the initial value of the magnetic field, $B_i$.
This initial state is then propagated in time. 
During the time period where the magnetic field is unchanged,
$B(t)=B_i$, the time propagation simply
modulates the phase of the initial state.
After the magnetic field changes to the value $B(t)=B_m$,
the system is projected onto the new eigenstates at that field.
We then propagate the quantum amplitudes of the two-atom energy eigenstates.  
At the end of the double pulse, when the magnetic
field becomes $B_f$, our final state is expanded 
in the $\phi_\nu$ states corresponding to the final value of the 
magnetic field, $B_f$. 
Using the correspondence between two-body and many-body states
discussed in the previous section,
we then relate the final population probabilities 
to the Donley {\it{et al.}} experiment.  In taking the absolute 
square of this final quantum amplitude for each distinguishable
final state, cross terms arise that exhibit quantum beats, the
most prominent of which is between the molecular state and the atomic 
trap ground state (condensate).
The parameters entering our simulations are the 
rescaled frequency $\omega'$, the
four magnetic field values 
$B_i$, $B_{evolve}$, $B_m$ and $B_f$, 
as well as the three time periods
$t_1$, $t_{evolve}$ and $t_2$. 
Our particular simulation values are given in the caption
of Fig. \ref{fig3}. 

\begin{figure}
\vspace{10 mm}
\begin{center}
\includegraphics[width=4in]{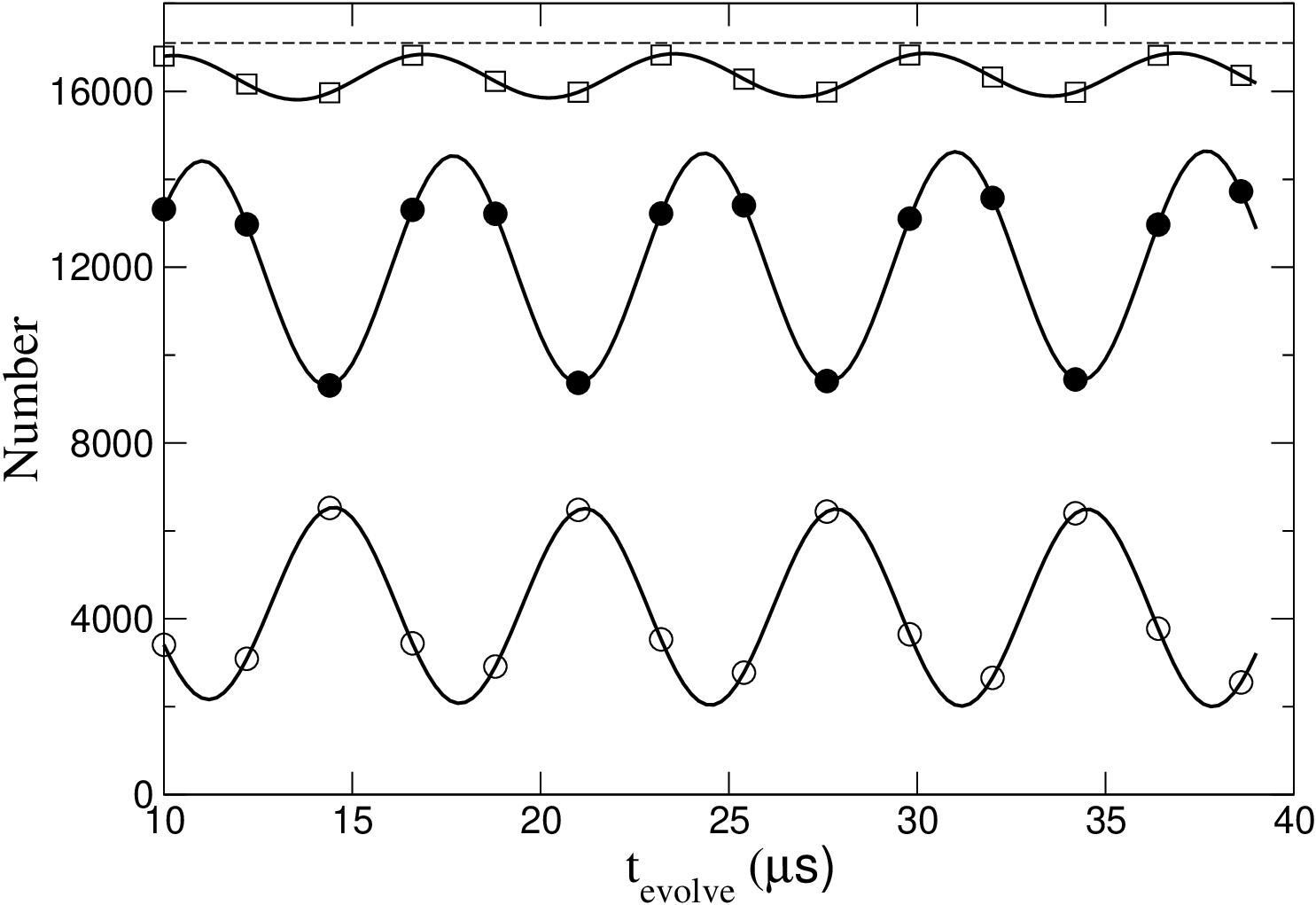}
\end{center}
\caption{\label{fig4}
Population of the ground trap state (filled circles)
and of the excited trap states (open circles) 
at the end of the magnetic field pulse shown
in Fig. \ref{fig4}. The sum of the two populations
is plotted with squares and allows determination 
of the number of atoms that made the transition to molecules
by subtraction from the total number of atoms ($N=17100$).
}
\end{figure}

Figure \ref{fig4} shows the results
of our time-dependent two-atom calculation for
a scaled frequency $\omega'$ that gives a two-atom density 
of $4\cdot10^{12}$ cm$^{-3}$.  This figure can be directly compared with 
Fig. 6 of Ref. \cite{Donley02}. 
Inspection of these two figures shows reasonable agreement
in the average values and the oscillation amplitudes for
the  three components. In addition, our results reproduce  approximately
the relative phases of the oscillations of the 
three different populations.  The observed agreement is
remarkable if one considers that we have used the sudden approximation
for magnetic field ramps that in the experimental conditions
last about $15$ $\mu$s. We have also assumed that the atomic
sample has a constant (average) density across the atom cloud, while an
improved model could consider averaging over a distribution
of densities (see, for example, \cite{Kokel02}). 
The frequency of the beats that we observe is approximately
150 kHz in accordance with the bound state energy 
produced by our parametrization of the scattering length
given in  Eq. (\ref{asc}).
In contrast, the experiment measures oscillations at approximately 
196 kHz for the same value of  $B_{\rm evolve}$.
In fact, a possible way to improve our model would be to use a better 
description of the atom-atom scattering properties, 
and consequently, a more accurate value of the molecular
binding energy. Our estimations show that 
inclusion of the energy dependence 
up to the effective range level (the first non-zero correction, see
\cite{Blume02}) significantly improves the agreement between 
the bound state energy produced by our simple model and by 
a multichannel calculation. However, in order to 
maintain our simple description of the system dynamics 
during the pulse evolution,
we do not include this correction here.

Note that our rescaled frequency yields
$\hbar \omega' /k_B \approx 180$ nK, which implies that 
the first excited pair above the trap ground state  would share
approximately two times this energy. We note that 
the resulting energy per atom for the least energetic 
of our ``hot atoms'' is approximately 180 nK above the 
trap ground state. This is not very far from
the experimentaly observed energy (150 nK) of the burst atoms. 
However, including all excited trap states,
our model predicts that burst atoms will have a
higher average energy  than in  the experiment.
In addition, our model predicts a dependence of the energy of the hot atoms 
on the initial atomic density like the
one given by Eq. (\ref{omegap}) while the experiment
does not mention an observable density dependence of the 
energy of the hot bursts. We  conclude that the rescaling 
used in our model allows  a reasonable prediction of the 
number of the hot atoms, but  has a limited ability
to account for  their energy  distribution.

\begin{figure}
\vspace{10 mm}
\begin{center}
\includegraphics[width=4in]{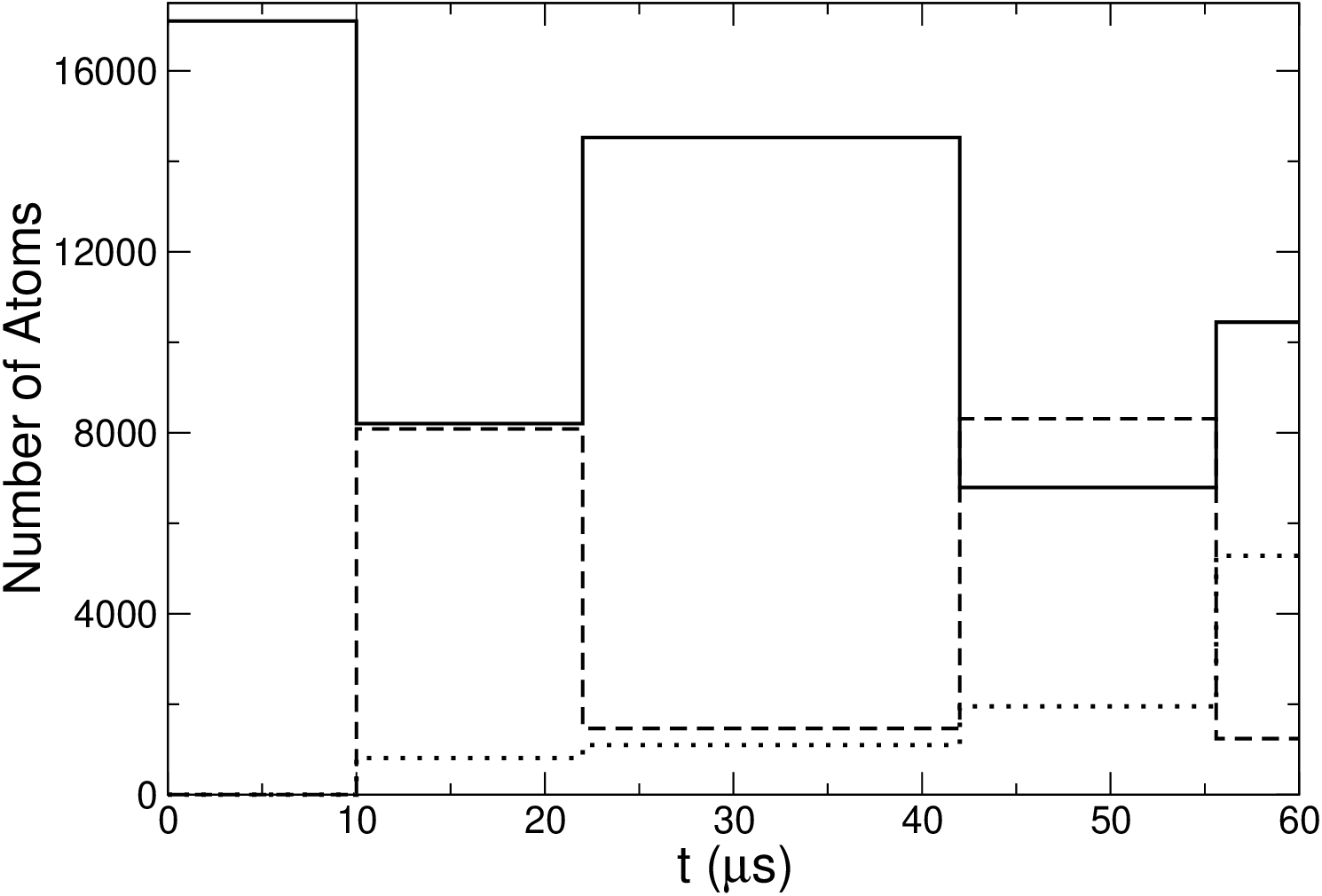}
\end{center}
\caption{\label{fig5}
Evolution of population of the three types of states
as a function of time during one magnetic field double-pulse 
(with $t_{\rm evolve}=20$ $\mu$s, 
identical to the one plotted in Fig. \ref{fig3}).
Solid line: trap ground state population (condensate);
dashed line: molecular state population;
dotted line: population of the excited trap states (``burst'' atoms).
}
\end{figure}

Figure \ref{fig5} shows the evolution of the 
populations during only one pulse (with $t_{\rm evolve}=20$ $\mu$s)
as a function of time. 
According to our calculations, the molecular eigenstate
and the excited states (i.e., the non-condensed atoms)
acquire roughly comparable populations after the 
complete first pulse (after $t=22$ $\mu$s)
and mantain them until the last ramp of the pulse, when 
the population of the excited
states becomes considerably larger. 
The significant population we observe for the molecular 
state during $t_{\rm evolve}$
supports our interpretation of the quantum beats seen in the 
end-of-pulse populations.  These beats are the result
of the interference of quantum 
paths that go through the intermediate molecular state with those that go
through the intermediate ground or excited trap states. (The difference
between the latter is too small to show up 
on the time scales considered here.)
The predictions of this two-atom model, 
concerning the intermediate time populations
of the molecular and hot components, are consistent with the comments
of Braaten {\it et al.}, \cite{Braaten03} (see also \cite{Stoof02})
 which point out the difficulty of interpreting the intermediate populations 
shown in Ref. \cite{Kokel02}.  Interpretation of the results of
Kokkelmans and Holland \cite{Kokel02} is complicated by the fact that
their two-body representation does not consist of eigenfunctions of the
molecular Hamiltonian.  Accordingly, their "molecular state" is not 
actually the two-body molecular eigenstate, except at magnetic fields well
above 160 G.  In problems with a linear Schr\"odinger equation, a simple
basis change could always be carried out, to re-express the physics in
an eigenrepresentation.  Here, however, the nonlinearity of the coupled
equations complicates this transformation. 
At the same time, as Braaten {\it et al.} \cite{Braaten03} comment, the 
success of the final calculations in reproducing the experimental
observations with no adjustable parameters is immediately 
apparent and convincing
that the right two-body physics has been incorporated into the formulation.
We mainly recommend caution in interpreting the meaning of 
the ``molecular state" in the Kokkelmans and Holland
formulation, except at high magnetic fields where it approximately
coincides with a two-body eigenstate.

The two-atom model disagrees with a specific qualitative prediction
of Mackie {\it et al.} \cite{Mackie02}. Whereas in our model
the ``hot'' atoms at the end of the pulse can be created 
through any of the three intermediate states (i.e, molecules,
condensed atoms or non-condensed atoms), in the model of Ref.
\cite{Mackie02} the ``hot'' atoms are solely the results 
of the ``rogue dissociation'' of intermediate molecular states.
Our approach suggests that around half 
of the final hot atoms are not produced  by rogue dissociation.
In fact, the pathway {\it condensate} $\rightarrow$ {\it hot atoms}, 
which is apparently neglected by Ref. \cite{Mackie02},
is of comparable importance.
This may in fact be the additional loss mechanism that is cited as
being ``missing'' in Ref. \cite{Mackie02}.

\section{Conclusions}

We have investigated the results given by a simple
model of the dynamics of two trapped 
atoms near a Feshbach resonance.
Our model accounts for the interaction between the
two atoms using a zero range potential and it also includes the confinement
of the atoms by an external harmonic trap.
This model can be used to make predictions 
regarding  atomic Bose condensate driven with the help of 
magnetic fields near a Feshbach resonance, if
a rescaled frequency is used to achieve a density
comparable to the density of the many-atom condensates studied
experimentally.
A correspondence can be drawn between the 
three components observed in the recent
field-ramp experiments
and three groups of two-body states in our model.  
The experimentally-observed burst of
hot atoms appears in our model as atom pairs excited 
to states energetically higher than the trap
ground state level. The two-atom model is able to predict the populations
of these three states, in fair agreement with the experimental
observations. Accordingly this may provide a useful alternative view 
of the physics of coherent atom-molecule coupling in a condensate.  
The success of a two-body description 
may initially seem surprising, because the dominant physical
processes occurring in condensate experiments near a
Feshbach resonance are normally viewed 
as being inherently many-body in nature.  
Nevertheless, the present study suggests that a two-body 
picture, with minimal modifications,
is sufficiently realistic to be used for simple 
estimates at a qualitative or semiquantitative level.

\ack

We thank E. Cornell for helpful suggestions and encouragement.  
We also thank N. Claussen, J. Dunn, M. Holland, 
and C. Wieman for informative discussions.
This work was supported by NSF.

\end{document}